\documentclass[prd,a4paper]{revtex4}
\usepackage{epsfig}
\newcommand{\be}{\begin{equation}}
\newcommand{\ee}{\end{equation}}
\newcommand{\bea}{\begin{eqnarray}}
\newcommand{\eea}{\end{eqnarray}}
\newcommand{\bean}{\begin{eqnarray*}}
\newcommand{\eean}{\end{eqnarray*}}

\newcommand{\gapproxeq}{\lower
.7ex\hbox{$\;\stackrel{\textstyle >}{\sim}\;$}}
\newcommand{\lapproxeq}{\lower
.7ex\hbox{$\;\stackrel{\textstyle <}{\sim}\;$}}

\begin{document}

\bibliographystyle{unsrt}

\title{\bf Search for tetraquark candidate $Z(4430)$ in meson photoproduction}

\author{Xiao-Hai Liu$^1$ and Qiang Zhao$^{1,2}$}

\affiliation{1) Institute of High Energy Physics, Chinese Academy of
Sciences, Beijing 100049, P.R. China}

\affiliation{2) Department of Physics, University of Surrey,
Guildford, GU2 7XH, United Kingdom}

\author{Frank E. Close}

\affiliation{Rudolf Peierls Centre for Theoretical Physics,
University of Oxford, Keble Rd., Oxford, OX1 3NP, United Kingdom}

\date{\today}

\begin{abstract}

We propose a search for the newly discovered tetraquark candidate
$Z(4430)$ in photoproduction.  Based on the Belle results we show
that if $Z(4430)$ is a genuine resonance, its significantly large
coupling to $\psi^\prime\pi$ will cause it to stand out above the
background in  $\gamma p\to Z^+(4430) n\to \psi^\prime \pi^+ n$. We
consider the dependence of the cross section for the quantum numbers
($J^{P}=1^{-}$, $1^{+}$ or $0^{-}$).

\end{abstract}

\maketitle



\section{Introduction}

A resonant structure in the invariant mass of $\psi^\prime\pi^+$ at
4.43 GeV with a width of 45 MeV was recently reported by Belle in
$B\to K \psi^\prime \pi$~\cite{belle-z4430}.  This result
immediately provoked much interest from experiment and theory due to
the possibility of it being a tetraquark
candidate~\cite{maiani-2005a,maiani-2005b,maiani,rosner-2005,bigi-2005}
with flavor $c\bar{c}u\bar{d}$ or $c\bar{c}d\bar{u}$, i.e. a
charmonium with isospin $I=1$.

It has also been proposed that this structure may be due to
subthreshold effects in $D^*\bar{D_1}$
rescattering~\cite{bugg,rosner,rosner-2006,close-hadron07,meng-chao}.
Similar to the $X(3872)$ which is proposed to be a possible bound
state of $D\bar{D^*}$ due to long-range $\pi$ exchange
forces~\cite{tornqvist,cpage, swanson}, such a mechanism will favor
$J^{P}=0^{-}, 1^{-}$ or $2^{-}$ for $Z(4430)$ with $I=1$
~\cite{close-hadron07}. Also, it has been suggested by Maiani {\it
et al.}~\cite{maiani} that the strong coupling of the $Z(4430)$ to
$\psi^\prime\pi$ instead of $J/\psi\pi$ could be a signal for a
radially excited $1^{+-}$ tetraquark. Long ago Close and
Lipkin~\cite{close-lipkin-78} proposed that $c\bar{c} + \pi$ is a
channel that could identify tetraquark states such as $1^{+-}$ in
S-wave or, most interesting, $1^{--}$ in P-wave, hence we shall also
consider that $Z(4430)\to\psi^\prime\pi$ occurs via $P$-wave
transition if it has $J^{PC}=1^{--}$. In particular, a large
coupling to $\psi^\prime\pi$ will have special implications for its
photoproduction rate, and information about its quantum numbers may
be extracted.

Seeing this state in an independent production process will be
essential for establishing its existence and determining its nature.
We propose here to look for the $Z(4430)$ in photoproduction. If
$Z(4430)$ is a genuine signal for a tetraquark state which couples
to $\psi^\prime\pi$, then it can necessarily be photoproduced via
$\pi$ exchange through the $\gamma-\psi^\prime$ intermediate
coupling.  It suggests that $\gamma p\to Z^+(4430) n\to \psi^\prime
\pi^+ n$ could be a dominant production mechanism for $Z(4430)$.
Therefore, by looking for the signal of $Z(4430)$ in $\gamma p\to
Z^+(4430) n\to \psi^\prime \pi^+ n$, one may establish this state as
a genuine resonance. We also estimate a possible destructive
interference from the intermediate $\gamma-J/\psi$ coupling.
Although Belle does not see the decay of $Z(4430)\to J/\psi\pi$,
which implies a weaker $Z J/\psi\pi$ coupling, the relatively larger
leptonic width for $J/\psi\to e^+ e^-$ than $\psi^\prime \to e^+e^-$
may still lead to sizeable contributions from intermediate $J/\psi$
in photoproduction. We will investigate such effects in a vector
meson dominance (VMD) model.

In addition, we consider background contributions to this channel
which involve Pomeron exchanges and nucleon Born terms.
Interestingly, this shows that these backgrounds have rather small
interferences with the $Z^+(4430)$ production. This enhances the
possibility of identifying or refuting this state in
photoproduction.
 Experimental data from DESY may be able to clarify
this directly.

In the following  we first provide a detailed description of the
theoretical model by considering $Z(4430)$ to have either
 $J^{P}=1^{-}$,
$1^{+}$ or $0^{-}$. The model predictions for the differential and
total cross sections will be presented. We then make a quantitative
analysis of the possible background contributions to $\psi^\prime
\pi N$ final state. Dalitz plot analysis will be provided.

\section{The Model}

\begin{figure}[tb]
\begin{center}
\begin{tabular}{ccc}
\includegraphics[scale=1.0]{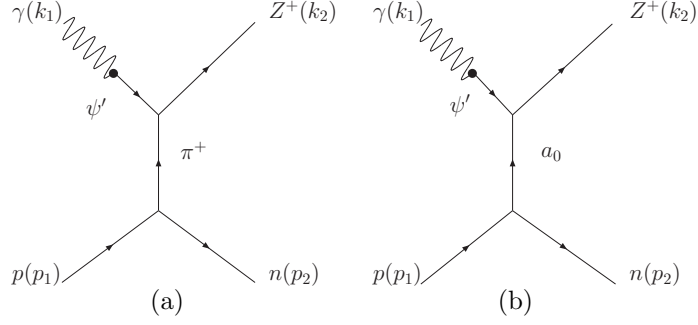}
\end{tabular}
\caption{$Z$ production through meson exchange} \label{fig:1}
\end{center}
\end{figure}

The production of $Z^+$ in $\gamma p\to Z^+ n$ can be described by
the processes in Fig.~\ref{fig:1}. Its production possesses
different features at different kinematics. We outline those
features below as they are essential for constructing the
theoretical model:

I) In the $t$-channel, as the electromagnetic (EM) interaction does
not conserve isospin, the $Z(4430)$ can be produced by either
isoscalar or isovector exchange. Taking into account the flavor
content, i.e. $Z^+$ has a hidden $c\bar{c}$ which does not carry
isospin, the isoscalar exchange in the $t$-channel will be forbidden
for the charged $Z(4430)$ while suppressed for the neutral partner.
We will come back to this point later.

Furthermore, as the $Z(4430)$ has only been seen in the
$\psi^\prime\pi$ channel, we shall concentrate on the $t$-channel
pion exchange as a leading contribution. We shall argue that this is
likely to give a lower limit for the photoproduction cross section.
We note in advance that the inclusion of e.g. $a_0$ exchange will
increase the cross sections since there is no interferences between
the pion and $a_0$ exchange amplitudes. We will present some
numerical estimates for the $a_0(980)$ exchange later. Heavier
$t$-channel meson (Regge) exchanges contribute to the cross section
at off-forward angles and could become important in the higher
energy region. As we do not have information about their couplings
to $Z(4430)\psi^\prime$, we shall focus on the near-threshold region
where the dominant pion exchange can be well-constrained and
reliably estimated.

II) For $J^{P}=1^{-}$ and $0^{-}$, $Z(4430)$ will couple to $VP$ in
$P$-wave at leading order, while for $J^{P}=1^{+}$, it is via
$S$-wave. We apply different effective Lagrangians to describe the
$Z\psi^\prime\pi$ vertex.

III) There is no available information about the coupling for a $Z$
to nucleons. This will appear in the $s$ or $u$-channel where
$Z(4430)$ can directly couple to $p$ and $n$. The $s$-channel
process generally becomes important near threshold, and at middle
and backward scattering angles. With increasing energies, these
contributions die out quickly due to the suppression of the $s$ and
$u$-channel propagators and the effects of the nucleon's form
factors. In contrast, the $t$-channel pion exchange may remain
dominant at forward angles and give the major contribution to the
cross section.

Based on the above rather general considerations, we identify
processes that will drive the $Z(4430)$ production. As the $Z(4430)$
is observed to couple strongly to $\psi^\prime \pi$, we propose to
look for signals in $\gamma p\to Z^+(4430) n \to \psi^\prime \pi^+
n$, and $\gamma p\to Z^0(4430) p\to \psi^\prime \pi^0 p$. This gives
rise to some background contributions as shown in Fig.~\ref{fig:2},
in particular, in the neutral $Z(4430)$ production.

IV) As the $Z(4430)$ is observed to couple to $\psi^\prime\pi$, we
derive the $Z\gamma \pi$ coupling by Vector Meson Dominance (VMD),
and assume that the $Z\gamma \pi$ coupling is due to a sum of
intermediate vector mesons which connect the photon with the
$Z(4430)$ and the exchanged $\pi$. We initially assume that
$\psi^\prime$ is the dominant intermediate vector meson in the
$Z\gamma\pi$ coupling, and then consider the possible effect of a
coupling to $J/\psi \pi$ .

\subsection{Production of $Z(4430)$ with $J^{P}=1^{-}$}

For $J^{P}=1^{-}$, the following Lagrangian is applied for the
$Z\psi^\prime \pi$ coupling, e.g. for the charged $Z(4430)$,
\begin{equation}\label{lagrangian-v}
{\cal
L}_{Z\psi^\prime\pi}=\frac{g_{Z\psi^\prime\pi}}{M_Z}\varepsilon_{\mu\nu\alpha\beta}
\partial^{\mu}\psi^{\prime\nu} \partial^\alpha Z^{+\beta} \pi^{-} +
H.c. \ ,
\end{equation}
where $\varepsilon_{\mu\nu\alpha\beta}$ is Levi-Civita tensor, and
the coupling constant $g_{Z\psi^\prime\pi}$ will be determined by
the width $\Gamma_{Z^\pm \to \psi^\prime
\pi^\pm}$~\cite{belle-z4430}, i.e.
\begin{equation}\label{gamma-z}
\Gamma_{Z\to  \psi^\prime\pi} = \frac{1}{8\pi} \frac{|{\cal
M}|^2}{3} \frac{|{\bf p}_{\pi cm}|}{M_Z^2} \ ,
\end{equation}
where
\begin{eqnarray}
|{\cal M}|^2 &=& \left(\frac{g_{Z\psi^\prime\pi}}{M_Z}\right)^2
[2(p_Z\cdot p_{\psi^\prime})^2 - 2p_Z^2 p_{\psi^\prime}^2] \nonumber \\
&=&\left(\frac{g_{Z\psi^\prime\pi}}{M_Z}\right)^2 \left[
\frac{(M_Z^2+M_{\psi^\prime}^2-m_\pi^2)^2}{2} - 2M_Z^2
M_{\psi^\prime}^2 \right] \ ,
\end{eqnarray}
and ${\bf p}_{\pi cm}$ is the three-vector momentum of the pion in
the $Z$ meson rest frame. With $ M_Z=4433$ MeV, and $\Gamma_{Z\to
\psi^\prime\pi} = 45$ MeV, we have $g_{Z\psi^\prime\pi}/M_Z= 2.365\
\mbox{GeV}^{-1}$. Any coupling to $J/\psi \pi$ will be analogously
related to $g_{Z J/\psi\pi}/M_Z$, which in turn is driven by the (as
yet unobserved) decay ${Z\to J/\psi\pi}$.

For the meson coupling vertices, we apply a form factor as follows:
\begin{equation}
F_Z = \frac{M_{\psi^\prime}^2 - m_i^2}{M_{\psi^\prime}^2 - q^2}
\end{equation}
with $m_i=m_\pi$, and the cutoff is set as the mass of the
intermediate vector meson~\cite{friman}, i.e. $\Lambda =
M_{\psi^\prime}$.

The general expression for an intermediate-vector-meson $V$ coupling
to $\gamma$ at leading order is
\begin{equation}
{\cal L}_{V\gamma} = \frac{eM_V^2}{f_V}V_\mu A^\mu
\end{equation}
the coupling constant $e/f_V$ can be determined by $V\to
e^{+}e^{-}$,
\begin{equation}\label{vmd}
\frac{e}{f_V} = \left[\frac{3\Gamma_{V\to e^+ e^-}}{2\alpha_e
|p_e|}\right]^{1/2} \ ,
\end{equation}
where $|p_e|$ is the electron three-moment in the vector meson rest
frame, and $\alpha_e = 1/137$ is the EM fine-structure constant.
With $\Gamma_{\psi^\prime\to e^+e^-}=2.48\pm 0.06$
keV~\cite{pdg2006}, it gives $e/f_{\psi^\prime}= 0.0166$.

The $\gamma$-$J/\psi$ coupling analogously is $e/f_{J/\psi}= 0.027$.
As this is larger than for the $\psi^\prime$ it is possible that any
coupling $Z \to J/\psi\pi$ could cause an additional contribution to
the photoproduction amplitude, which could destructively interfere
with the former and hence give a much reduced cross section. If this
were the case, it would also imply that a branching ratio $Z \to
J/\psi \pi$ should be visible. We shall discuss the dependence of
the photoproduction cross section on this.

At the meson nucleon coupling vertices we adopt the commonly used
Lagrangian
\begin{eqnarray}
{\cal L}_{\pi NN}^{int} &=& -i g_{\pi NN} \bar{N}\gamma_5
(\vec{\tau}\cdot\vec{\pi}) N  \nonumber \\
&=& -ig_{\pi NN}(\bar{p}\gamma_5 p \pi^0 +\sqrt{2}\bar{p}\gamma_5 n
\pi^{+} + \sqrt{2}\bar{n}\gamma_5 p\pi^{-} - \bar{n}\gamma_5 n\pi^0)
\ ,
\end{eqnarray}
where a standard value, $g_{\pi NN}^2/4\pi=14$, is
adopted~\cite{goldberger-treiman}. In addition, a form factor is
applied for the $\pi NN$ vertex,
\begin{equation}
F_{\pi NN} = \frac{\Lambda_\pi^2 - m_\pi^2}{\Lambda_\pi^2 - q^2} ,
\end{equation}
with $\Lambda_\pi =0.7$ GeV. We note that this form factor
successfully accounts for the photoproduction of $\omega$ and
$\rho$ meson near threshold~\cite{friman,kirchbach}. Some works
have argued for a larger value~\cite{ericson}. Such values would
increase the cross sections relative to what we compute here, and
so we use the cited value in the spirit of seeking a lower limit
for the cross-section.

The effective Lagrangian leads to the amplitude for producing a
charged $Z(4430)$ via charged pion exchange in $\gamma p \to
Z^+(4430) n$:
\begin{eqnarray}\label{trans-v1}
T_{1fi} &=& -i\left(\sqrt{2}g_{\pi NN} \frac{g_{Z\psi^\prime\pi
}}{M_Z} \frac{e}{f_{\psi^\prime}}\right) \bar{u}(p_2)\gamma_5 u(p_1)
\varepsilon_{\mu\nu\alpha\beta}k_1^\mu k_2^\alpha
\epsilon^\nu \epsilon_Z^{*\beta} \nonumber \\
&& \times\frac{1}{q^2-m_\pi^2} F_{\pi NN}(q^2)
F_{Z\psi^\prime\pi}(q^2) \ ,
\end{eqnarray}
where $p_1$ ($k_1$) and $p_2$ ($k_2$) are four-vector momenta of
initial nucleon (photon), and final nucleon ($Z$ meson),
respectively; $\epsilon^\mu$ and $\epsilon_Z^{*\beta}$ are the
polarization vectors of $\gamma$ and $Z$, respectively.

By defining $t \equiv (p_1 - p_2)^2 \equiv q^2$, $s\equiv
(k_1+p_1)^2$, the differential cross section is
\begin{equation}
 \frac{d\sigma}{dt} = \frac{1}{64\pi s} \frac{1}{|k_{1cm}|^2}
 \frac{1}{4} |{\cal M}|^2  \ ,
\end{equation}
where the invariant matrix element squared gives
\begin{eqnarray}\label{trans-v-total}
|{\cal M}|^2 &=& \sum\limits_{pol} \left| T_{1fi}  \right|^2 \nonumber \\
&=&\left(\sqrt{2}g_{\pi NN} \frac{g_{Z\psi^\prime\pi }}{M_Z}
\frac{e}{f_{\psi^\prime}}\right)^2
\frac{-q^2(q^2-M_Z^2)^2}{(q^2-m_\pi^2)^2} \left(\frac{\Lambda_\pi^2
-m_\pi^2}{\Lambda_\pi^2-q^2} \right)^2 \left(\frac{M_{\psi^\prime}^2
-m_\pi^2}{M_{\psi^\prime}^2-q^2} \right)^2  \ .
\end{eqnarray}

After integrating over the range of $|t|$, i.e. within $t_{max}$ and
$t_{min}$,
\begin{equation}
t_{max}(t_{min}) = \frac{M_Z^4}{4s} - (k_{1cm}\mp k_{2cm})^2
\end{equation}
the total cross section can be obtained. In the above equation,
$k_{1cm}= \frac{E_\gamma M_N}{\sqrt{s}}$ is the photon energy in the
overall c.m. system, where $E_\gamma$ is the photon energy in the
rest-frame of the initial proton.

In the Appendix, we provide the analogous expressions for
$t$-channel $a_0$ exchange. Note that the pion and $a_0$ exchanges
contribute to the real and imaginary part of the transition
amplitude respectively. Thus, there are no interferences between
these two channels, and the inclusion of the $a_0$ exchange will
enhance the $Z(4430)$ cross section. As we lack knowledge on the
$Z\psi^\prime a_0$ and $a_0 NN$ coupling strengths, which are in any
event expected to be much smaller than the $\pi$ exchange
contributions, we will focus on the pion exchange here.

For neutral $Z$ production contributions may also be allowed from
Pomeron exchange. However, we argue that this scenario is suppressed
due to the following reasons: i) In the case where the
 photon (isoscalar component) couples to $c\bar{c}$, isospin has
to be transferred to the pair of light $q\bar{q}$ in order to form
$Z^0(4430)$. ii) The photon can couple to isovector vector mesons
made of light $q\bar{q}$, which then pull out a pair of $c\bar{c}$
after exchanging a Pomeron. Although this process conserves isospin,
it is suppressed by the soft creation of a large mass $c\bar{c}$,
and also as it requires the $Z$ meson to be formed with the
$q\bar{q}$ and $c\bar{c}$ pairs at relatively large momentum
transfers. Consequently suppression from the final state
wavefunction is likely. Precise estimates will be model dependent
but for the purpose of estimating a lower limit for the production
cross section, we neglect this contribution here.

In contrast, Pomeron contributions can generate background. As shown
in Fig.~\ref{fig:2}(a) and (b), the direct production of
$\psi^\prime$ allows Pomeron exchanges as an important production
mechanism. An intermediate nucleon will then decay into a pion and
nucleon. In principle, all the intermediate isospin 1/2 nucleon
resonances can contribute to this process. Again, as a simple
estimate of the background, we only consider the nucleon pole
contributions.

\subsection{Production of $Z(4430)$ with $J^{P}=1^{+}$}

For the production of $Z(4430)$ with $J^{P}=1^{+}$, the difference
from $J^{PC}=1^{-}$ is at the $Z\psi^\prime\pi$  vertex for which
the following effective Lagrangian is adopted~\cite{xiong, haglin}:
\begin{eqnarray}
\mathcal {L}_{Z\psi^\prime\pi} &=& \frac{g_{Z\psi^\prime\pi}}{M_Z}
\left( \partial^\alpha \psi^{\prime\beta} \partial_\alpha \pi
Z_\beta -
\partial^\alpha \psi^{\prime\beta} \partial_\beta \pi Z_\alpha \right)  \ ,
\end{eqnarray}
where the notations are the same as Eq.~(\ref{lagrangian-v}).
Schematically, the $\pi$ exchange transition can be illustrated by
Fig.~\ref{fig:1}, and the definition of kinematic parameters is the
same as the previous subsection.

The transition matrix elements are as follows:
\begin{eqnarray}
T_{1fi} &=& -i\left(\sqrt{2}g_{\pi NN}
\frac{g_{Z\psi^\prime\pi}}{M_Z} \frac{e}{f_{\psi^\prime}}\right)
\bar{u}(p_2)\gamma_5 u(p_1) \epsilon_Z^{*\mu} \epsilon^\nu
[k_1\cdot(k_2-k_1)g_{\mu\nu}-k_{1\mu}(k_2-k_1)_\nu]  \nonumber \\
& \times & \frac{1}{q^2-m_\pi^2} F_{\pi NN}(q^2)
F_{Z\psi^\prime\pi}(q^2)  \ ,
\end{eqnarray}
which gives
\begin{eqnarray}\label{trans-v-axi}
|{\cal M}|^2 &=& \sum\limits_{pol} \left| T_{1fi}  \right|^2 \nonumber \\
&=&\left(\sqrt{2}g_{\pi NN} \frac{g_{Z\psi^\prime\pi }}{M_Z}
\frac{e}{f_{\psi^\prime}}\right)^2
\frac{-q^2(q^2-M_Z^2)^2}{(q^2-m_\pi^2)^2} \left(\frac{\Lambda_\pi^2
-m_\pi^2}{\Lambda_\pi^2-q^2} \right)^2 \left(\frac{M_{\psi^\prime}^2
-m_\pi^2}{M_{\psi^\prime}^2-q^2} \right)^2  \ .
\end{eqnarray}
One notices that the above equations have the same form as
Eq.~(\ref{trans-v-total}). This is because the longitudinal terms in
$Z\gamma\pi$ coupling vanish in the real photon limit for the chiral
partner $1^{-}$ and $1^{+}$.

The difference between these two parities leads to different values
for the decay constant $g_{Z\psi^\prime\pi}$. With the partial
width:
\begin{eqnarray}
\Gamma_{Z^+ \to \psi^\prime \pi^+}
=\left(\frac{g_{Z\psi^\prime\pi}}{M_Z}\right)^2 \frac{ |{\bf p}_{\pi
cm}|}{24\pi M_Z^2}
 \left[ 2(p_\pi \cdot p_{\psi^\prime})^2 +
M_{\psi^\prime}^2(m_\pi^2+|{\bf p}_{\pi cm}|^2)\right] \ ,
\end{eqnarray}
we have
\begin{equation}
g_{Z\psi^\prime\pi}/M_Z=2.01\ \mbox{GeV}^{-1} \ ,
\end{equation}
which is smaller than that for $1^{-}$.

\subsection{Production of $Z(4430)$ with $J^{P}=0^{-}$}

If we take $Z(4430)$ as a pseudoscalar meson with $J^P=0^-$, the
following effective lagrangian for $Z\psi^\prime \pi$ coupling is
adopted:
\begin{equation}
\mathcal
{L}_{Z\psi^\prime\pi}=ig_{Z\psi^\prime\pi}(\pi^-\partial_\mu  Z^+
-\partial_\mu \pi^- Z^+)\psi^{\prime \mu}
\end{equation}
Similarly, the $\pi$ exchange transition can be illustrated in
Fig.~\ref{fig:1}(a). The transition matrix element is as follows:
\begin{eqnarray}
T_{fi} &=& i\left(\sqrt{2}g_{\pi NN} g_{Z\psi^\prime\pi}
\frac{e}{f_{\psi^\prime}}\right) \bar{u}(p_2)\gamma_5 u(p_1)
\frac{(q+k_2)\cdot \epsilon(k_1)}{q^2-m_\pi^2} F_{\pi NN}(q^2)
F_{Z\psi^\prime\pi}(q^2)
\end{eqnarray}
which gives
\begin{eqnarray}\label{tran-p}
|{\cal M}|^2 &=& \sum\limits_{pol} \left| T_{fi} \right|^2 \nonumber \\
&=&\left(\sqrt{2}g_{\pi NN} \frac{g_{Z\psi^\prime\pi}}{M_Z}
\frac{e}{f_{\psi^\prime}}\right)^2 \frac{-8q^2
M_Z^4}{(q^2-m_\pi^2)^2} \left(\frac{\Lambda_\pi^2
-m_\pi^2}{\Lambda_\pi^2-q^2} \right)^2 \left(\frac{M_{\psi^\prime}^2
-m_\pi^2}{M_{\psi^\prime}^2-q^2} \right)^2.
\end{eqnarray}
Again, the coupling coupling constant $g_{Z\psi^\prime\pi}$ is also
determined by $\Gamma_{Z^\pm\to \psi^\prime \pi^\pm}$:
\begin{equation}
\Gamma_{Z\to \psi^\prime
\pi}=\left(\frac{g_{Z\psi^\prime\pi}}{M_Z}\right)^2 \frac{|{\bf
p}_{\pi cm}|}{6\pi} \left[ \frac{(p_{\psi^\prime}\cdot
p_\pi)^2}{M_{\psi^\prime}^2} - p_\pi^2 \right],
\end{equation}
and we have
\begin{equation}
g_{Z\psi^\prime\pi}/M_Z=1.39\ \mbox{GeV}^{-1} \ .
\end{equation}

Interestingly, the neutral $Z^0\to \psi^\prime \pi^0$ has
$J^{PC}=0^{--}$, which is an exotic quantum number.

\subsection{Numerical results for $\gamma p\to Z^+ n$ }

\begin{figure}[tb]
\begin{center}
\begin{tabular}{ccc}
\includegraphics[scale=1.0]{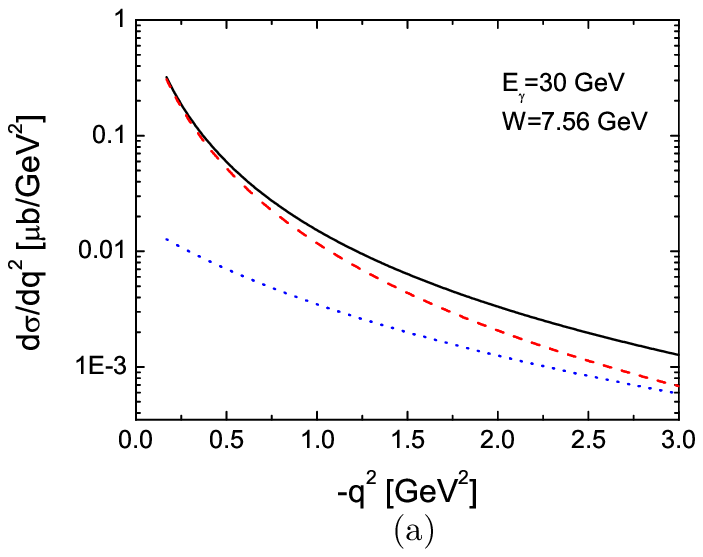}
\includegraphics[scale=1.0]{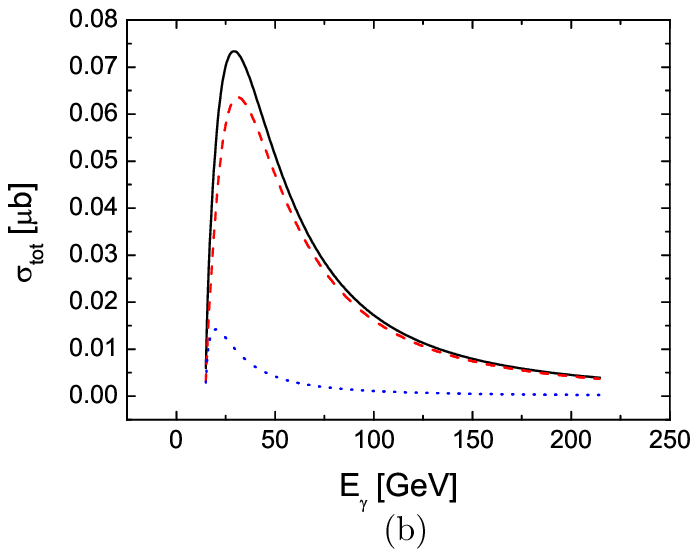}
\end{tabular}
\caption{Differential and total cross sections for $\gamma p\to Z^+
n$ where the spin-parity of $Z$ is $J^{P}=1^{-}$. The dashed and
dotted lines denote the $\pi$ and $a_0$-exchange contributions,
respectively, and the solid line is the sum of both.}
\label{fig:vec}
\end{center}
\end{figure}

\begin{figure}[tb]
\begin{center}
\begin{tabular}{ccc}
\includegraphics[scale=1.0]{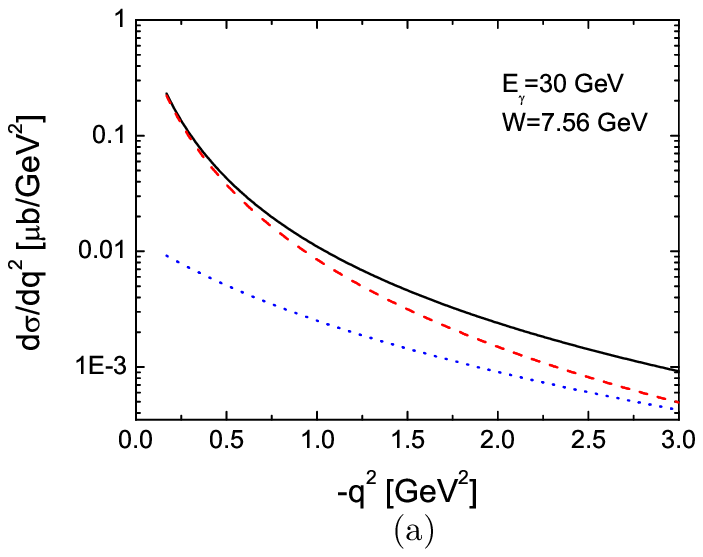}
\includegraphics[scale=1.0]{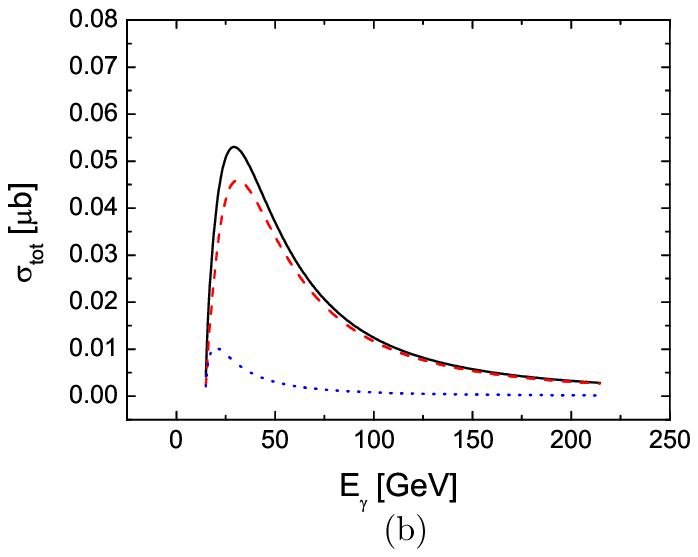}
\end{tabular}
\caption{Differential and total cross sections for $\gamma p\to Z^+
n$ where the spin-parity of $Z$ is $J^{P}=1^{+}$. The notations are
the same as Fig.~\protect\ref{fig:vec}.} \label{fig:axi}
\end{center}
\end{figure}

\begin{figure}[tb]
\begin{center}
\begin{tabular}{ccc}
\includegraphics[scale=1.0]{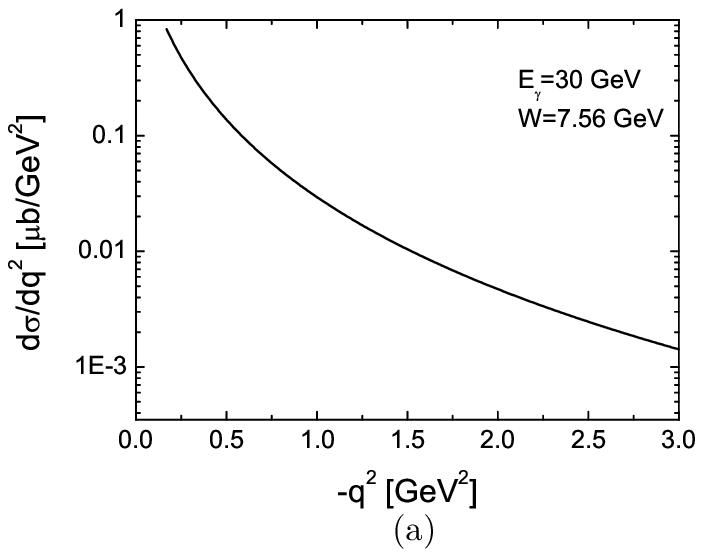}
\includegraphics[scale=1.0]{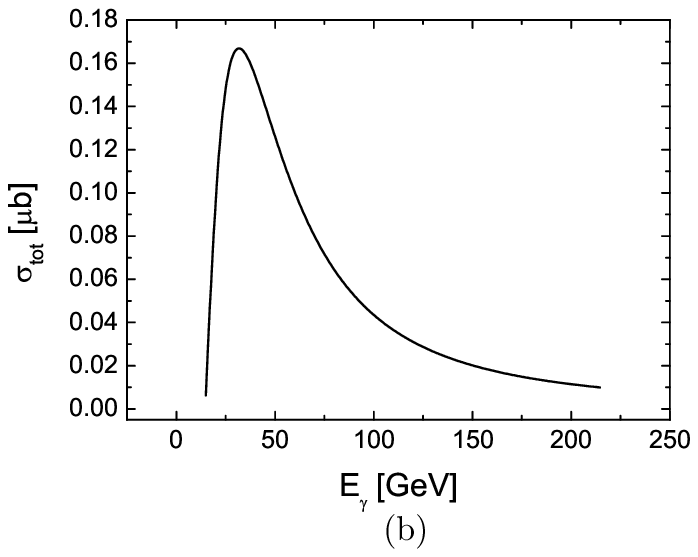}
\end{tabular}
\caption{Differential and total cross section for $\gamma p\to Z^+
n$ through $\pi$-exchange where the spin-parity of  $Z$ is
$J^P=0^-$. } \label{fig:pse}
\end{center}
\end{figure}

In Figs.~\ref{fig:vec}, \ref{fig:axi} and \ref{fig:pse}, the
differential and total cross sections are plotted for $J^{P}=1^{-}$,
$1^{+}$, and $0^-$, respectively. The photon energy for the
differential cross sections is $E_\gamma=30$ GeV ($W=7.56$ GeV),
which corresponds to the peaking energy region in the total cross
sections. In all three cases, the forward peaking turns to be a
prominent feature in the differential cross sections due to the pion
exchange.

The results of Figs.~\ref{fig:vec}, \ref{fig:axi} and \ref{fig:pse}
show compatible production cross sections for those three
spin-parity assignments. For spin-parity $1^-$ and $1^+$, their
production cross sections have the same expression in the real
photon limit as noted before. The difference between
Figs.~\ref{fig:vec} and \ref{fig:axi} arises from the different
coupling strengths for $g_{Z\psi^\prime\pi}$ extracted from $Z^+\to
\psi^\prime\pi^+$.

The cross section for $0^-$ production is found larger than the
other two assignments. This is understandable by comparing
Eqs.~(\ref{trans-v-total}), (\ref{trans-v-axi}), and (\ref{tran-p})
with each other. It also shows that in order to determine the
quantum numbers, measurement of the angular distributions of $Z\to
\psi^\prime\pi$ is necessary. For spin-parity $1^+$, the decay will
be via relative $S$-wave between $\psi^\prime\pi$, while it is via
relative $P$-wave for $1^-$ and $0^-$. Therefore, an observation of
$P$-wave decay will require further studies such as polarization
observables to determine the quantum numbers of $Z(4430)$.

For $0^-$, $a_0$ exchange is forbidden and the pion exchange is the
exclusive leading contribution. For other $J^P$ $a_0$ could
contribute but in practice need not concern us here. Even were its
coupling as large as that of the pion, $g_{Z\psi^\prime
a_0}=g_{Z\psi^\prime \pi}$, its contribution to the cross-section is
negligible (in Figs.~\ref{fig:vec} and \ref{fig:axi}, the results
due to $t$-channel $a_0$ exchanges are presented by the dotted
curves). This in practise allows us to focus on the pion exchange
mechanism in the photoproduction reaction as a reliable estimate of
the lower bound of the production cross sections.

Due to the pion exchange the energy dependence of the total cross
sections exhibits a strong threshold enhancement in $\gamma p \to
Z^+ n$. It is similar to the charged $\rho$ meson production near
threshold where the charged pion exchange (unnatural parity
exchange) also plays a dominant
role~\cite{bauer-78,zhao-li-bennhold-98}. However, such a threshold
enhancement seems to be absent in charmonium production, such as
$J/\psi$ and $\psi^\prime$~\cite{hera-2000,H1}. This may be due to
the dominance of the diffractive process which submerges the
contributions from $t$-channel unnatural parity exchanges. Although
$J/\psi$ has a relatively large coupling to $\rho\pi$, whereby
$t$-channel pion exchange turns to be important in $\gamma p\to
J/\psi p$, the coupling is not as large as that for
$Z\psi^\prime\pi$. Consequently, the predicted threshold enhancement
in $Z$ meson photoproduction is strongly driven by the large
$Z\psi^\prime\pi$ coupling.

Our main concern therefore is the possibility of some further
contribution that is large and destructive. The most important
question here seems to concern the role played by the intermediate
$J/\psi$ in the VMD model, which in a worst case could destructively
interfere with the $\psi^\prime$. By defining
\begin{equation}
\frac{\Gamma_{Z\to J/\psi\pi}}{\Gamma_{Z \to \psi^\prime \pi}}
\equiv x \ ,
\end{equation}
we include $J/\psi$ contributions to the $\pi$ exchange diagrams;
i.e. the real photon couples to $J/\psi$ and $\psi^\prime$, which
then couple to $Z\pi$. We then have
\begin{equation}
T_{fi} = T_{fi}^{\psi^\prime}+T_{fi}^{J/\psi} = \left(1 +
\frac{g_{ZJ/\psi\pi}}{g_{Z\psi^\prime\pi}}
\frac{f_{\psi^\prime}}{f_{J/\psi}} e^{i\theta} \right)
T_{fi}^{\psi^\prime}
\end{equation}
where we have already set the form factors
\begin{equation}
F_{ZJ/\psi \pi}(q^2) = F_{Z\psi^\prime\pi}(q^2)
\end{equation}
for simplicity, and $e^{i\theta}$ is the relative phase between the
two transition matrix element. Couplings $g_{ZJ/\psi\pi}$ and
$f_{J/\psi}$ can be determined by Eq.~(\ref{gamma-z}) and
Eq.~(\ref{vmd}), respectively. The cross section becomes
\begin{equation}
\sigma^{J/\psi+\psi^\prime}_{\gamma p\to Z^+ n} \simeq
(1+0.75^2x+1.5\sqrt{x} \cos{\theta})\sigma^{\psi^\prime}_{\gamma
p\to Z^+ n} \ .
\end{equation}
The worst destructive situation is at $\theta=\pi$. But the
destructive effects will depend on the value of $x$. Note that Belle
has not seen $Z\to J/\psi\pi$ decay, which allows an estimate of
$x\leq 0.1$, and with which the total cross section would be still
sizeable.

If the $Z$ meson is a genuine resonance, the threshold enhancement
will be a signature for its existence. This can be further clarified
by the study of background effects in $\gamma p\to \psi^\prime \pi^+
n$ in the next section.

\section{Background analysis}

\begin{figure}[tb]
\begin{center}
\begin{tabular}{ccc}
\includegraphics[scale=1.0]{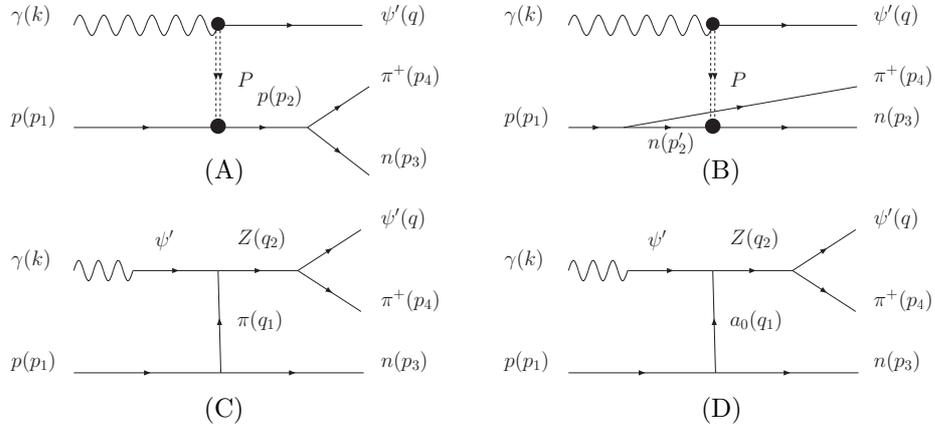}
\end{tabular}
\caption{Major processes that contribute to the background in
$\gamma p\to \psi^\prime\pi^+ n$. } \label{fig:2}
\end{center}
\end{figure}

Experimentally, the final state particles identified are
$\psi^\prime$, $\pi$ and nucleon. As a result, background
contributions to the  $\psi^\prime\pi N$ channel can interfere with
the transitions we are interested in. As discussed earlier, one of
the major contributions is due to the diffractive production of
$\psi^\prime$ (Fig.~\ref{fig:2}). In this section, we combine
$\gamma p\to Z^+ n\to \psi^\prime \pi^+ n$ and $\gamma p\to
\psi^\prime p\to \psi^\prime \pi^+ n$ together to make an estimate
of the signal and background contributions. This could be useful for
experimental search for the $Z^+$ in this channel. Since the
production rate for all these three spin-parities are similar, we
only consider $J^P=1^-$ in the following analysis.

The schematic transition diagrams are illustrated in
Fig.~\ref{fig:2}. We first extend the formulae of the previous
section to $\gamma p\to Z^+ n\to \psi^\prime \pi^+ n$, and then
include the diffractive contributions from the Pomeron exchanges.

\subsection{Meson exchange terms}

By including the vertex coupling for $Z^+\to \psi^\prime\pi^+$, the
transition amplitude for pion exchange can be explicitly written
down
 \begin{eqnarray}
 T^{\mathcal{M}}_{1fi} &=& \sum\limits_{pol} \Gamma_{\gamma 
\psi^\prime}\Gamma_{Z\psi^\prime\pi}^1\Gamma_{Z\psi^\prime\pi}^2
\Gamma_{pn\pi} \times  F_{Z\psi^\prime\pi}(q_1^2)F_{\pi
 NN}(q_1^2)F_{Z\psi^\prime\pi}(q_2^2)\nonumber \\
&=& -i\left[ \sqrt{2}g_{\pi NN} \frac{e}{f_\psi^\prime} \left(
\frac{g_{Z\psi^\prime\pi}}{M_Z} \right)^2 \right] \bar{u}(p_3)
\gamma_5 u(p_1) \epsilon_\mu(k)\epsilon_{\psi^\prime}^{\nu}(q)
\varepsilon_{\mu_1\nu_1\alpha_1\beta_1}k^{\mu_1}q_2^{\alpha_1}
\varepsilon_{\mu_2\nu\alpha_2\beta_2}q^{\mu_2}q_2^{\alpha_2}\nonumber \\
&\times&
\frac{g^{\mu\nu_1}g^{\beta_1\beta_2}}{(q_1^2-m_\pi^2)(q_2^2-M_Z^2+iM_Z\Gamma)}
F_{Z\psi^\prime\pi}(q_1^2)F_{\pi
 NN}(q_1^2)F_{Z\psi^\prime\pi}(q_2^2) \ ,
 \end{eqnarray}
where the vertex functions are
 \begin{eqnarray}
 \Gamma_{\gamma \psi^\prime} &=&
 \frac{eM_{\psi^\prime}^2}{f_{\psi^\prime}}\epsilon_\mu(k)\epsilon_{\psi^\prime}^{\mu}(k)\\
 \Gamma_{Z\psi^\prime\pi}^1 &=&
 \frac{g_{Z\psi^\prime\pi}}{M_Z}\varepsilon_{\mu_1\nu_1\alpha_1\beta_1}k^{\mu_1}\epsilon_{\psi^\prime}
^{\nu_1}(k)
 q_2^{\alpha_1}\epsilon_Z^{\beta_1}(q_2) \\
\Gamma_{Z\psi^\prime\pi}^2 &=&
 \frac{g_{Z\psi^\prime\pi}}{M_Z}\varepsilon_{\mu_2\nu\alpha_2\beta_2}q^{\mu_2}\epsilon_{\psi^\prime}^{
\nu}(q)
 q_2^{\alpha_2}\epsilon_Z^{\beta_2}(q_2) \\
 \Gamma_{pn\pi} &=& -i\sqrt{2}g_{\pi NN}\bar{u}(p_3)\gamma_5
 u(p_1) \ ,
 \end{eqnarray}
 and the form factor is
 \begin{equation}
F_{Z\psi^\prime\pi}(q_2^2) =\frac{\Lambda_Z^2-M_Z^2}{\Lambda_Z^2 -
q_2^2}.
 \end{equation}
Considering a moderate modification of the cutoff will not change
the results significantly, we take $\Lambda_Z=M_{\psi^\prime}$ as an
input value. The kinematic variables are denoted in
Fig.~\ref{fig:2}.


\subsection{Pomeron exchange term}

Pomeron as a phenomenology accounting for the diffractive
transitions has been widely studied in the
literature~\cite{donnachie,pichowsky,laget}. In this approach, the
Pomeron mediates the long range interaction between a confined quark
and a nucleon, and has been shown to behave just like a $C=+1$
isoscalar photon.

The Pomeron-nucleon coupling is determined by the vertex:
\begin{equation}
F_\mu(t)=3\beta_0\gamma_\mu f(t)
\end{equation}
where $t$ is the Pomeron momentum squared, $\beta_0$ represents the
coupling constants between a single Pomeron and a light constituent
quark, and $f(t)$ is the isoscalar nucleon electromagnetic form
factor:
\begin{equation}
f(t)=\frac{4M_N^2-2.8t}{(4M_N^2-t)(1-t/0.7)^2} \ .
\end{equation}
For the $\gamma V \mathcal{P}$ vertex ($V$ represents the
corresponding vector meson, here is $\psi^\prime$), we take the
on-shell approximation for restoring gauge invariance
as~\cite{zhaoq}. Therefore the equivalent vertex for the $\gamma
\psi^\prime \mathcal{P}$ is reduced to a loop tensor as follows:
\begin{equation}
\frac{2\beta_c \times 4\mu_0^2}{(M_{\psi^\prime}^2
-t)(2\mu_0^2+M_{\psi^\prime}^2 -t)} T^{\mu\alpha\nu} \ ,
\end{equation}
where
\begin{equation}
T^{\mu\alpha\nu} = (k+q)^\alpha g^{\mu\nu} -2k^\nu g^{\alpha\mu}
\end{equation}
and $\beta_c$ is the effective coupling constant between Pomeron and
a charm quark within $\psi^\prime$; $\mu_0$ is a cutoff of the form
factor related to the Pomeron.

The amplitudes for Fig.~\ref{fig:2}(a) and (b) can then be expressed
respectively as
\begin{eqnarray}
T^{\mathcal{P}}_{1fi} &=& 24\sqrt{2}\beta_{0}\beta_{c}g_{\pi NN}
\frac{\mu_0^2
f(t)\mathcal{G}_P(s,t)}{(M_{\psi_\prime}^2-t)(2\mu_0^2+M_{\psi^\prime}^2-t)}
T^{\mu\alpha\nu}\epsilon_{\psi^\prime \nu}(q) \epsilon_{\mu}(k)
\bar{u}(p_3) \gamma_{5} \frac{p\!\!\!/_2 + M_N}{p_2^2 - M_N^2}
\gamma_{\alpha} u(p_1) \nonumber \\
&& \\
T^{\mathcal{P}}_{2fi} &=& 24\sqrt{2}\beta_{0}\beta_{c}g_{\pi NN}
\frac{\mu_0^2
f(t)\mathcal{G}_P(s,t)}{(M_{\psi^\prime}^2-t)(2\mu_0^2+M_{\psi^\prime}^2-t)}
T^{\mu\alpha\nu}\epsilon_{\psi^\prime \nu}(q) \epsilon_{\mu}(k)
\bar{u}(p_3) \gamma_{\alpha} \frac{p\!\!\!/_2^\prime +
M_N}{p_2^{\prime 2} - M_N^2} \gamma_{5} u(p_1) \ .
\end{eqnarray}
Function $\mathcal{G}_P(s,t)$ is related to the Pomeron trajectory
$\alpha(t)=1+\epsilon+\alpha^\prime t$ via
\begin{equation}
\mathcal{G}_P(s,t)= -i(\alpha^\prime s)^{\alpha(t)-1} \ .
\end{equation}

We also consider a form factor for the vertex $pn\pi$ since the
intermediate nucleon is off-shell, i.e.
\begin{eqnarray}
F_{pn\pi}(s_{34}) &=&
\frac{\Lambda_{pn\pi}^2-M_N^2}{\Lambda_{pn\pi}^2 - s_{34}} \\
F_{pn\pi}(u_{14}) &=&
\frac{\Lambda_{pn\pi}^2-M_N^2}{\Lambda_{pn\pi}^2 - u_{14}}
\end{eqnarray}
where $s_{34} = (p_3 + p_4)^2$, $u_{14}=(p_1 - p_4)^2$. Considering
the kinematic region of the three-particles final state, we take a
cutoff $\Lambda_{pn\pi}=1.0\ \mbox{GeV}$ in our following
calculation to avoid the singularity.

\subsection{Numerical results with background contributions}

Combining the signal terms and background amplitudes, the invariant
transition amplitude becomes
\begin{eqnarray}
|{\cal M}|^2 &=& \sum\limits_{pol} \left|T_{1fi}^\mathcal{P} +
T_{2fi}^\mathcal{P}
+ T_{1fi}^\mathcal{M} + T_{2fi}^\mathcal{M}\right|^2 \nonumber \\
&\equiv& |A+B+C+D|^2 \nonumber \\
&=& |A|^2 + |B|^2 + |C|^2 + |D|^2 \nonumber \\
&+& 2Re[AB^*] + 2Re[AC^*] + 2Re[AD^*]+2Re[BC^*]+2Re[BD^*]+2Re[CD^*]
\ .
\end{eqnarray}
where $2Re[CD^*]=0$, and $2Re[AC^*]$, $2Re[AD^*]$, $2Re[BC^*]$,
$2Re[BD^*]$ are found to be much smaller than the other terms.

The following values are applied in the Pomeron exchange amplitudes:
\begin{eqnarray}
 && \beta_0^2=4.0 \ \mbox{GeV}^2, \ \beta_c^2=0.592\ \mbox{GeV}^2 \nonumber \\
 && \alpha^\prime=0.25\ \mbox{GeV}^{-2}, \ \epsilon=0.08, \ \mu_0=1.2\
 \mbox{GeV} \ ,
\end{eqnarray}
where $\beta_c$ is determined by considering the ratio
$\sigma_{\gamma p \to \psi^\prime p}/{\sigma_{\gamma p \to \psi
p}} $ through the Pomeron exchange. We take the following value
H1~\cite{H1} as a constraint:
\begin{equation}
\frac{\sigma_{\gamma p \to \psi^\prime p}}{\sigma_{\gamma p \to
\psi p}} =0.15 \ .
\end{equation}
With the coupling $\beta_{c,J/\psi }^2=0.8\ \mbox{GeV}^2$ fixed by
the $J/\psi$ photoproduction~\cite{laget}, the coupling $\beta_c$
for $\psi^\prime$ is determined. We note that these parameters
lead to a reasonable description of both $J/\psi$ and
$\psi^\prime$ photoproduction in comparison with the
experiment~\cite{h1-data}.

As shown by the total production cross section for $\gamma p\to Z^+
n$, the largest cross section is at $W\simeq 8\ \mbox{GeV}(
E_\gamma\simeq 34\ \mbox{GeV})$. We thus take this energy as an
input to simulate the invariant mass spectrum for $\gamma p\to
\psi^\prime\pi^+ n$. This could be the kinematic region which favors
the separation of the $Z^+$ signals.

\begin{figure}
  \begin{center}
  \includegraphics[scale=1.0]{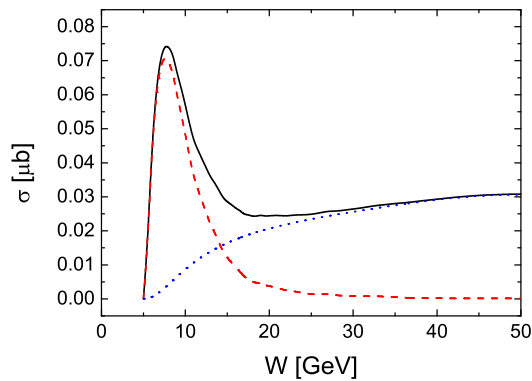}
  \caption{Energy dependence of total cross section for $\gamma p \to
\psi^\prime \pi^+ n$. $W$ is the c.m. energy. The dashed line
denotes the meson ($\pi$ and $a_0$) exchange contribution, the
dotted line is the Pomeron-exchange contribution, and the solid line
is the full contribution. }\label{fig:xsbg}
  \end{center}
\end{figure}

In Fig.~\ref{fig:xsbg}, we calculate the total cross sections for
$\gamma p\to \psi^\prime \pi^+ n$ including both signal and
background contributions. As shown by the solid curve, the overall
production cross section has a quick increase just above threshold
and is dominated by the $Z^+$ production as shown by the dotted
curve. In contrast, the background contributions from the Pomeron
exchange are relatively small and do not vary drastically with the
increasing energies at $W > 20$ GeV. This is consistent with the
familiar diffractive behavior in vector meson photoproduction.

The steep rise of the cross section near threshold turns out to be
very different from other known heavy vector photoproductions, such
as $\phi$, $J/\psi$ and $\psi^\prime$. In those cases, the threshold
cross sections are rather smooth and no threshold enhancement has
been observed. In the production of $Z^+(4430)$, its strong coupling
to $\psi^\prime\pi$ gives rise to the predominant contributions from
pion exchange. Note that $\psi^\prime\to e^+ e^-$ is also strong.
These lead to an unusual prediction for its cross section near
threshold.

\begin{figure}[htbp]
\begin{center}
  \includegraphics[scale=1.0]{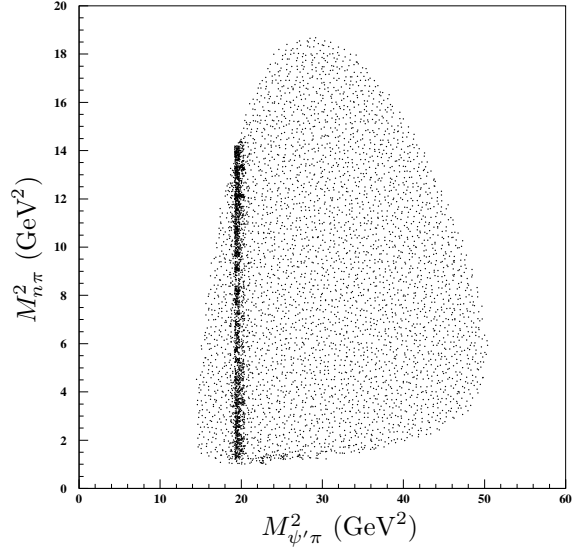}
\caption{Dalitz plot of $\gamma p \to n \psi^\prime \pi^+$ at $W=8$
GeV ($E_\gamma\simeq 34$ GeV). The axial variables are
$M_{\psi^\prime \pi}^2=(q+p_4)^2$, and $M_{n\pi}^2=(p_3+p_4)^2$. }
\label{fig:dalitz1}
\end{center}
\end{figure}

It is more constructive to inspect the Dalitz plot for the
three-body final state. We take a sample of the events at $W=8$
and $12$ GeV, respectively. As shown by Fig.~\ref{fig:dalitz1}, a
clear band appears in the invariant mass spectrum of
$\psi^\prime\pi$ due to the $Z$ resonance. Although we do not
include the $s$ and $u$ channel nucleon resonances, their
contributions seem not to demolish the $Z(4430)$ band since their
effects will add to the plot horizontally and in the low
$M^2_{n\pi}$ region, e.g. $M^2_{n\pi}< 5$ GeV$^2$.

\begin{figure}[htbp]
  \begin{center}
  \includegraphics[scale=1.0]{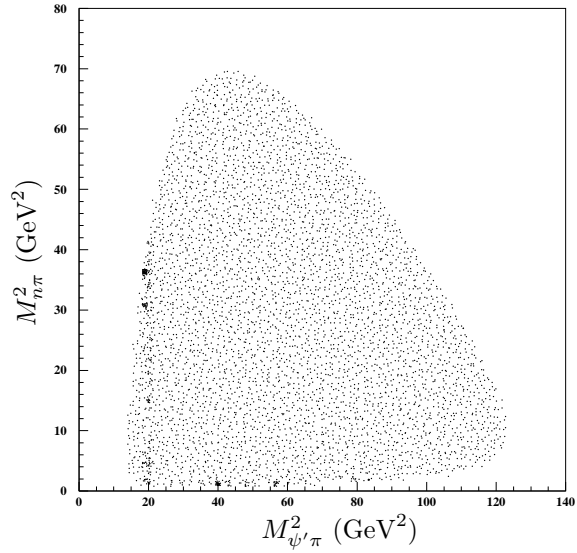}
\caption{Dalitz plot of $\gamma p \to n \psi^\prime \pi^+$ at $W=12$
GeV ($E_\gamma\simeq 76$ GeV). The axial variables are the same as
Fig.~\protect\ref{fig:dalitz1}. } \label{fig:dalitz2}
  \end{center}
\end{figure}

With the increasing center-mass energy, the background from
Pomeron-exchange becomes important, and the $Z(4430)$ signals are
submerged as shown by Fig.~\ref{fig:dalitz2}.

\begin{figure}
  \begin{center}
  \includegraphics[scale=1.0]{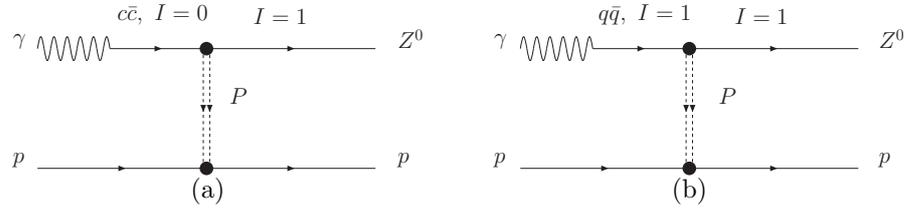}
\caption{Schematic diagrams for the diffractive production of
$Z^0(4430)$ in $\gamma p\to Z^0 p$. Figure (a) and (b) illustrate
subprocesses for producing $Z^0$ via intermediate $I=0$ and $I=1$
vector meson components in the VMD model. }\label{fig-z0}
  \end{center}
\end{figure}

For the neutral $Z^0(4430)$ production, the leading background
contribution is the same as the charged $Z$ production. In the
$t$-channel $Z^0$ production, an additional transition is the
diffractive production of $Z^0$ via Pomeron exchange for $1^{-}$ or
odderon exchange for $1^+$. It gives rise to two possibilities as
shown in Fig.~\ref{fig-z0}. In Fig.~\ref{fig-z0}(a), the initial
photon couples to the isoscalar component of a vector meson
($c\bar{c}$), which then couples to $Z^0$ by projecting the
$c\bar{c}$ component to the $I=1$ tetraquark wavefunction. Note that
the exchange of Pomeron or odderon does not compensate the isospin.
This process will be suppressed due to isospin violation.  The other
diffractive transition is that the initial photon couples to the
isovector component of the intermediate meson as shown by
Fig.~\ref{fig-z0}(b). For the tetraquark scenario, the isovector
component must be from the light $q\bar{q}$ instead of $c\bar{c}$,
hence requiring creation of a pair of $c\bar{c}$ from the vacuum
which then couples to the $I=1$ tetraquark wavefunction. Note that
the $c\bar{c}$ pair is much heavier than the light $q\bar{q}$. To
form a tetraquark of $cq\bar{c}\bar{q}$, it needs a sizeable overlap
between the final state $Z^0$ and the intermediate isovector
$q\bar{q}$. Since the $c\bar{c}$ is much heavier than the light
$q\bar{q}$, the $Z^0$ wavefunction overlap with the $c\bar{c}$ and
light $q\bar{q}$ should be small. This allows us to neglect the
Pomeron or odderon exchange contributions to the direct $Z^0$
production via diffractive processes. Due to the above argument, the
neutral $Z^0$ photoproduction turns out to be similar to the charged
$Z$ production studied earlier.

The neutral $Z^0$ of $0^-$ assignment leads to production of exotic
quantum number $0^{--}$. If indeed the $Z^0(4430)\to\psi^\prime
\pi^0$ has a partial width of $\sim 45$ MeV, its photoproduction
cross section near threshold will also be significant. By measuring
the angular distribution of $Z^0\to\psi^\prime \pi^0$, its quantum
numbers can also be determined.

\section{Conclusion}

In this work we have made a detailed analysis of the photoproduction
reaction $\gamma p\to Z^+(4430) n$ in an effective Lagrangian
approach for possible assignments of $J^{P}=1^{-}$, $1^{+}$ and
$0^-$ to the $Z^0(4430)$. As suggested by the observation of Belle,
we show that the $t$-channel pion exchange is an important
production mechanism for creating $Z^+(4430)$ at the rate of several
tens of nano-barn. In this process the $\pi NN$ coupling can be well
constrained and the $Z\gamma\pi$ coupling can be extracted from the
$Z\psi^\prime\pi$ data via VMD model.

We have also checked the possible effect of $t$-channel $a_0$
exchange. We find its contributions are relatively small even if, as
an upper limit, we fix the $Z\psi^\prime a_0$ coupling to be the
same as $g_{Z\psi^\prime\pi}$. Since there is no interference
between the $a_0$ and pion exchange amplitudes, we can neglect the
$a_0$ exchange in the calculation and only adopt the relatively
well-defined pion exchange for estimating the lower bound of the
$Z(4430)$ photoproduction cross sections. We find that due to pion
exchange, a clear threshold enhancement is predicted for $Z(4430)$
photoproduction, which is different from photoproduction of $J/\psi$
and $\psi^\prime$ near threshold, and could be a signature for the
$Z(4430)$ existence. This is also the kinematic region that
uncertainties arising from other unknown processes can be overlooked
as a reasonable approximate.

We also consider possible background contributions to the signals
in $\gamma p\to \psi^\prime\pi^+ n$, where the diffractive
production of $\psi^\prime$ turns to be the leading source.
Interestingly, it shows that the interferences from the background
terms are rather small. In the Dalitz plot at $W=8$ GeV
($E_\gamma\simeq 34$ GeV), a clear band of the $Z$ meson signals
appear in the invariant mass spectrum of $\psi^\prime \pi$.
Similar results are obtained for the neutral production channel.
For other background contributions from diffractive transitions,
we argue that they may be suppressed by isospin conservation and
momentum mismatching. Although nucleon resonance contributions to
the background in Fig.~\ref{fig:2}(A) and (B) have not been
included in this calculation, their effects are unlikely to
submerge the $Z(4430)$ band in the Dalitz plot. The reason is that
the nucleon resonance cross sections will accumulate horizontally
in a form of uneven bands, and be located at $M^2_{n\pi} < 5$
GeV$^2$. In the kinematics of higher $M^2_{n\pi}$ interferences
from the nucleon resonances are expected to be negligible.  As a
consequence, the $Z(4430)$ production signals can be rather
cleanly isolated from the background, and the measurement of the
angular distributions between $\psi^\prime$ and $\pi$ will tell
whether they are in a relative $S$ or $P$ wave. The quantum
numbers of the $Z(4430)$ can thus be determined.

In summary, in order to establish the $Z(4430)$ as a concrete
candidate of tetraquark state and determine its quantum numbers,
further search for its signals in other production processes is
necessary. Supposing it is a genuine resonance with sizeable
partial decay width to $\psi^\prime\pi$~\cite{belle-z4430}, we
have argued that photoproduction should be useful for providing a
clear and consistent evidence for its existence. Experimental data
which have been collected at H1 and ZEUS may already contain such
information.

\section*{Acknowledgement}

Useful discussion with ZhiQing Zhang about experiments at HERA is
acknowledged. This work is supported, in part, by the U.K. EPSRC
(Grant No. GR/S99433/01), National Natural Science Foundation of
China (Grant No.10675131), and Chinese Academy of Sciences
(KJCX3-SYW-N2). F.E.C. acknowledges support from the U.K. Science
and Technology Facilities Council (STFC) and the EU-RTN programme
contract No. MRTN-CT-2006-035 482: ``Flavianet".

\section*{Appendix}

We give below the details for $a_0$ exchange.

\subsection{$J^P=1^-$}

The Lagrangian for the $Z\psi^\prime a_0$ is:
\begin{equation}
{\cal L}_{Z\psi^\prime a_0} = \frac{g_{Z\psi^\prime a_0}}{M_Z}
(\partial^\alpha Z^\beta \partial_\alpha \psi^\prime_\beta -
\partial^\alpha Z^\beta \partial_\beta \psi^\prime_\alpha)a_0 \ ,
\end{equation}
where $g_{Z\psi^\prime a_0}$ is the coupling constant. At this
moment, no information about this coupling is available. A
conservative estimate is to set it the same as $g_{Z\psi^\prime\pi}$
as an upper limit.

The Lagrangian for $a_0NN$ interaction is
\begin{eqnarray}
{\cal L}_{a_0 NN}^{int} &=& g_{a_0 NN} \bar{N}
(\vec{\tau}\cdot\vec{a_0}) N \nonumber \\
&=& g_{a_0NN}(\bar{p}pa_0 + \sqrt{2}\bar{p}n a_0
+\sqrt{2}\bar{n}pa_0 - \bar{n}na_0) \ ,
\end{eqnarray}
where $g_{a_0NN}^2/4\pi = 1.075$ is adopted. A form factor is also
applied
\begin{equation}
F_{a_0 NN} = \frac{\Lambda_{a_0}^2 - m_{a_0}^2}{\Lambda_{a_0}^2 -
q^2} \ ,
\end{equation}
with $\Lambda_{a_0}= 2.0$ GeV~\cite{friman,kirchbach}.

The amplitude for $a_0$ exchange is:
\begin{eqnarray}
T_{2fi} &=& \left( \sqrt{2} g_{a_0NN}\frac{g_{Z\psi^\prime
a_0}}{M_{\psi^\prime}} \frac{e}{f_{\psi^\prime}} \right)
\bar{u}(p_2) u(p_1) k_2^\alpha (k_{1\alpha} g_{\mu\beta} -
k_{1\beta}
 g_{\mu\alpha})\epsilon_Z^{*\beta} \epsilon^\mu \nonumber \\
 && \times\frac{1}{q^2-m_{a_0}^2} F_{a_0NN}(q^2)F_{Z\psi^\prime
 a_0}(q^2)\ .
\end{eqnarray}

\subsection{$J^P=1^+$}

For the production of $Z(4430)$ with $J^{P}=1^{+}$, its difference
from $J^{P}=1^{-}$ is at the $Z\psi^\prime\pi$  and $Z\psi^\prime
a_0$ vertices for which the following effective lagrangians are
adopted~\cite{xiong, haglin}:
\begin{eqnarray}
\mathcal {L}_{Z\psi^\prime\pi} &=& \frac{g_{Z\psi^\prime\pi}}{M_Z}
\left( \partial^\alpha \psi^{\prime\beta} \partial_\alpha \pi
Z_\beta -
\partial^\alpha \psi^{\prime\beta} \partial_\beta \pi Z_\alpha \right) \ , \\
{\cal L}_{Z\psi^\prime a_0} &=& \frac{g_{Z\psi^\prime
a_0}}{M_Z}\varepsilon_{\mu\nu\alpha\beta}
\partial^{\mu}\psi^{\prime\nu} \partial^\alpha Z^{\beta} a_0 \ ,
\end{eqnarray}
where the notations are the same as Eq.~(\ref{lagrangian-v}).

The transition matrix elements are as follows:
\begin{eqnarray}
T_{2fi} &=& \left(\sqrt{2}g_{a_0 NN} \frac{g_{Z\psi^\prime
a_0}}{M_Z} \frac{e}{f_{\psi^\prime}}\right) \bar{u}(p_2) u(p_1)
\varepsilon_{\mu\nu\alpha\beta}k_1^\mu k_2^\alpha
\epsilon^\nu \epsilon_Z^{* \beta} \nonumber \\
&\times& \frac{1}{q^2-m_{a_0}^2} F_{a_0 NN}(q^2) F_{Z\psi^\prime
a_0}(q^2) \ .
\end{eqnarray}


\newpage

\end{document}